\documentclass[aps,pra,reprint]{revtex4-2}%
\usepackage{amsmath}
\usepackage{array}
\usepackage{amsfonts}
\usepackage{amssymb}
\usepackage{graphicx}
\usepackage{color}
\usepackage{dcolumn}
\usepackage{fullpage}
\usepackage{bm}
\usepackage{hyperref}
\usepackage{epstopdf}
\usepackage{cancel}

\begin{document}

\title{\textit{Zitterbewegung}, momentum and spin dynamics of electromagnetic waves in linear dielectric medium}
\author{Adam B. Cahaya}
\email{adam@sci.ui.ac.id}
\affiliation{Department of Physics, Faculty of Mathematics and Natural Sciences, Universitas Indonesia, Depok 16424, Indonesia} 

\begin{abstract}

The momentum of light in dielectric media has been a century-long controversy that continues to attract significant interest. In a linear dielectric medium with refractive index \textit{n}, the momentum is predicted to be smaller by a factor of \textit{n} according to Abraham, and larger by the same factor according to Minkowski. By studying the coupled dynamics of electromagnetic waves and dipoles in a dielectric medium, we show that the change in momentum of the dipole, expressed by the Lorentz force, corresponds to the Abraham momentum and is given by the expectation value of the spin-projected momentum vector.  
On the other hand, the Minkowski momentum is obtained as the magnitude of the spin-projected momentum vector from the energy-momentum dispersion relation derived by diagonalizing the coupled Hamiltonian and determines the direction of refraction in accordance with Snell's law. Our model also predicts a \textit{zitterbewegung}-like oscillation due to helicity mixing between left- and right-handed wave components, mediated by dipole oscillation. These internal wave dynamics may be observable via wavepacket motion or polarization-sensitive measurements.

\end{abstract}
\maketitle
\section{Introduction} 
Due to the opposing predictions by Abraham and Minkowski, the momentum of light in dielectric media has been a century-long debate \cite{Kemp2011,Barnett2010,PhysRevA.95.043804}. Since the wavelength in a dielectric medium with refractive index $n$ is shortened, Minkowski predicted that the wavevector $k=2\pi n/\lambda$ and momentum are larger:
\begin{equation}
p_M=\hbar k=\frac{nhc}{\lambda}.
\end{equation}
On the other hand, Abraham argued that the reduction in velocity $v=c/n$ should reduce the momentum:
\begin{equation}
p_A=\frac{\hbar\omega}{c^2}v=\frac{hc}{n\lambda}.
\end{equation}
While some experiments support one prediction \cite{doi:10.1098/rspa.1954.0043,doi:10.1098/rspa.1980.0035,doi:10.1098/rspa.1978.0072,PhysRevLett.24.156,PhysRevLett.94.170403,PhysRevLett.101.243601,Ashkin:86,PhysRevA.81.063830}, no particular setup can conclusively disprove the other \cite{Grzegorczyk2008,Buchanan2007,MANSURIPUR20101997}.

Theoretical discussions of the Abraham-Minkowski controversy often employ field theory approaches \cite{Baxter2010,Griffiths2012}, which are well suited for describing the wave properties of particles. These discussions tend to focus on momentum density \cite{Aguirregabiria2004,Babson2009,PhysRevA.44.3985}.  
However, to properly study the momentum of light, it is necessary to consider the particle nature of the wave, e.g., photon-electron elastic collisions in Compton scattering. In Compton scattering, the momentum of light is clearly determined from the change in electron momentum \cite{Cooper_1985}.

In this article, we focus on a linear dielectric medium modeled as a Drude material, where electrons are connected by springs. This model yields a linear refractive index for electromagnetic (EM) waves at low frequencies \cite{Griffiths_2017}. The Drude model has been successfully used to describe electric, heat, and optical transport properties in materials \cite{Li:01}. The theoretical predictions of the Drude model are in reasonable agreement with observed values for many materials \cite{Soukoulis} and have been applied to describe Hall effects \cite{PhysRevLett.99.206601,8471174}. In this work, we model light as a classical electromagnetic wave, and all dynamics are derived within the semiclassical framework.

The rest of this article is organized as follows. Section \ref{Sec_Drude}  introduces the Drude dipole model and derives the refractive index. Section \ref{Sec_Hamiltonian} formulates the Hamiltonian matrix describing electromagnetic wave dynamics and presents the helicity mixing between left- and right-handed components mediated by dipole coupling. Section \ref{Sec_Result} presents the main results, including the derivation of Minkowski and Abraham momenta, velocity operator, spin precession, and a \textit{zitterbewegung}-like oscillation. Section \ref{Sec_Discussion} discusses connections to previous models and possible experimental consequences. Section \ref{Sec_Conclusion} summarizes. The emergence of \textit{zitterbewegung}-like behavior from internal mode coupling in the medium highlights wave dynamics that resemble features of particle models, bridging wave and particle pictures. 

\

\section{Method}

\subsection{Drude dipole model} 
\label{Sec_Drude} 

The classical Drude model focuses on the dynamics of electrons under an electric field $\mathbf{E}$. Here we include the dynamics of the medium by modeling the positive charge as connected to the electron by a spring with natural oscillation frequency $\omega_0$, yielding coupled equations
\begin{equation}
m_\pm \frac{d^2 \mathbf{r}_\pm}{dt^2} = - m \omega_0^2 \left(\mathbf{r}_\pm - \mathbf{r}_\mp\right) \pm e \mathbf{E} \pm e \frac{d\mathbf{r}_\pm}{dt} \times \mathbf{B},
\end{equation}
where $m_-$ and $m_+ \gg m_-$ are the masses of negative and positive charges, respectively, and $m = (m_+^{-1} + m_-^{-1})^{-1}$ is the effective mass of the relative dipole motion. The first term on the right-hand side represents the internal restoring force of the dipole, with natural oscillation frequency $\omega_0$, and $\mathbf{B}$ is the magnetic field.

By defining the center of mass $\mathbf{R} = (m_+ \mathbf{r}_+ + m_- \mathbf{r}_-) / M$ and relative position $\mathbf{r}_e = \mathbf{r}_- - \mathbf{r}_+$, one obtains
\begin{align}
M \frac{d^2 \mathbf{R}}{dt^2} &= -e \frac{d \mathbf{r}_e}{dt} \times \mathbf{B}, \label{EqMomentum}\\
\frac{d^2 \mathbf{r}_e}{dt^2} &= -\omega_0^2 \mathbf{r}_e - \frac{e}{m} \mathbf{E} - \frac{e}{m} \frac{d \mathbf{R}}{dt} \times \mathbf{B}, \label{EqDrude}
\end{align}
where $M = m_+ + m_-$. Equation~\ref{EqMomentum} describes the center of mass motion influenced by the net Lorentz force on the dipole, and will be used to determine the momentum of EM wave in the medium.

Equation~\ref{EqDrude} is the Drude equation for an electron attached to a spring with natural oscillation frequency $\omega_0$. For $m \ll M$, $\dot{\mathbf{R}}$ can be neglected. Using the Fourier transform $f(x,t) = (2\pi)^{-1} \int e^{i(kx - \omega t)} f(\omega) d\omega$, the equation can be linearized and solved:
\begin{align}
\mathbf{r}_e &= \frac{-e}{m(\omega_0^2 - \omega^2)} \mathbf{E}. \label{Eqr}
\end{align}
Note that while the initial derivation uses a single dipole oscillator for conceptual clarity, all dynamical quantities in the model, such as energy, polarization, and velocity, are defined per unit volume. This is implemented by dividing by the representative volume \( V \) so that the resulting dynamics describe the averaged response of a polarization mode within the dielectric.

The relative permittivity, related to the electric polarization $\mathbf{P} = - e \mathbf{r}_e / V$, is then
\begin{align}
\varepsilon_r &= 1 + \frac{\partial \mathbf{P}}{\varepsilon_0 \partial \mathbf{E}} = 1 + \frac{\omega_p^2}{\omega_0^2 - \omega^2}, \label{Eqer}
\end{align}
where $\omega_p = \sqrt{e^2 / \varepsilon_0 m V}$ is the plasma frequency, $1/V$ is the density, and $\varepsilon_0$ is the vacuum permittivity. In the low-frequency regime, the system behaves as a linear medium with refractive index
\begin{equation}
n = \lim_{\omega \ll \omega_0} \sqrt{\varepsilon_r} = \sqrt{1 + \frac{\omega_p^2}{\omega_0^2}}.
\label{EqRefractIndex}
\end{equation}
The derivation of this refractive index using a classical oscillator model is well known and appears in standard textbooks \cite{Griffiths_2017}. Reference~\cite{Stenholm1986} discusses similar dipole dynamics in the context of laser cooling. Compared to that treatment, we neglect spatial derivatives of the electric and magnetic fields $\nabla \textbf{E}$ and $\nabla \textbf{B}$, and focus on homogeneous field interactions in momentum space.

\subsection{Equation of motion of light}  
\label{Sec_Hamiltonian}
Coupled equations of the dipole and the EM wave arise from the Drude equation and Maxwell's equations for current $\mathbf{j} = \dot{\mathbf{P}} = - e \dot{\mathbf{r}}_e / V$:
\begin{equation}
\frac{\partial}{\partial t} 
\left[
\begin{array}{c}
\mathbf{E}\\
c \mathbf{B}\\
\dot{\mathbf{r}}_e\\
\mathbf{r}_e
\end{array}
\right]
=
\left[
\begin{array}{cccc}
0 & \frac{i c}{\hbar} \hat{\mathbf{p}} \cdot \mathbf{s} & \frac{e}{\varepsilon_0 V} & 0 \\
-\frac{i c}{\hbar} \hat{\mathbf{p}} \cdot \mathbf{s} & 0 & 0 & 0 \\
-\frac{e}{m} & 0 & 0 & -\omega_0^2 \\
0 & 0 & 1 & 0
\end{array}
\right]
\left[
\begin{array}{c}
\mathbf{E}\\
c \mathbf{B}\\
\dot{\mathbf{r}}_e\\
\mathbf{r}_e
\end{array}
\right]. \label{EqMaxwellDrude}
\end{equation}
Here $\hat{\mathbf{p}} = -i \hbar \nabla$ and $\mathbf{s}$ is a Hermitian SO(3) rotation representation \cite{Yang2022} fulfilling the angular momentum commutation relations $[s_i, s_j] = i \sum_k \epsilon_{ijk} s_k$ for spin 1, with $s^2 = 2$, and $s_z = 0, \pm 1$.

Equation~\ref{EqMaxwellDrude} can be written in Dirac form:
\begin{align}
i \hbar \frac{\partial}{\partial t} \Psi = H \Psi = \left[
\begin{array}{cc}
H_c & H_{ac} \\
H_{ac}^t & H_a
\end{array}
\right] \Psi, \label{EqDirac}
\end{align}
where 
$
\Psi = \left[
\begin{array}{cccc}
\mathbf{c}_+ & \mathbf{c}_- & \mathbf{a} & \mathbf{a}^\dagger
\end{array}
\right]^t,
$
with $$\mathbf{c}_\pm = \sqrt{\frac{\varepsilon_0 V}{2 \hbar \omega_0}} \left( \mathbf{E} \pm i c \mathbf{B} \right)$$ representing the right-handed and left-handed wave functions of the EM wave \cite{Smith2007} and 
\[
\mathbf{a} = \sqrt{\frac{m \omega_0}{2 \hbar}} \left( \mathbf{r}_e + i \frac{\dot{\mathbf{r}}_e}{\omega_0} \right), \quad
\mathbf{a}^\dagger = \sqrt{\frac{m \omega_0}{2 \hbar}} \left( \mathbf{r}_e - i \frac{\dot{\mathbf{r}}_e}{\omega_0} \right).
\]
Here $\mathbf{a}$ and $\mathbf{a}^\dagger$, which represent classical oscillator modes without quantization, are defined in analogy to annihilation and creation operators of a quantum harmonic oscillator for mathematical convenience.

This construction is similar to recent formulations of photonic wave equations presented in Ref.~\cite{Partanen2024}, where electric and magnetic field components are combined into an extended wave function. In contrast, our model also incorporates dipole degrees of freedom within the same structure, enabling explicit modeling of light-matter coupling within a unified dynamical framework. This inclusion of both field and medium dynamics distinguishes our approach from purely photonic descriptions.

The uncoupled Hamiltonians for the EM wave and dipole are
\begin{align}
H_c &= \left[
\begin{array}{cc}
c \hat{\mathbf{p}} \cdot \mathbf{s} & 0 \\
0 & -c \hat{\mathbf{p}} \cdot \mathbf{s}
\end{array}
\right], \quad
H_a = \left[
\begin{array}{cc}
\hbar \omega_0 & 0 \\
0 & -\hbar \omega_0
\end{array}
\right]
\end{align}
and the coupling Hamiltonian is
\begin{align}
H_{ac} = \frac{\hbar \omega_p}{2} \left[
\begin{array}{cc}
1 & -1 \\
1 & -1
\end{array}
\right].
\end{align}
The $H_{ac}$ enables indirect coupling between right- and left-handed EM waves. Since $H$ is Hermitian, the Heisenberg equation can be used to find time derivatives: $df/dt = [f, H] / i \hbar$.

\section{Result}
\label{Sec_Result}
\begin{figure}[tb]
\includegraphics[width=\columnwidth]{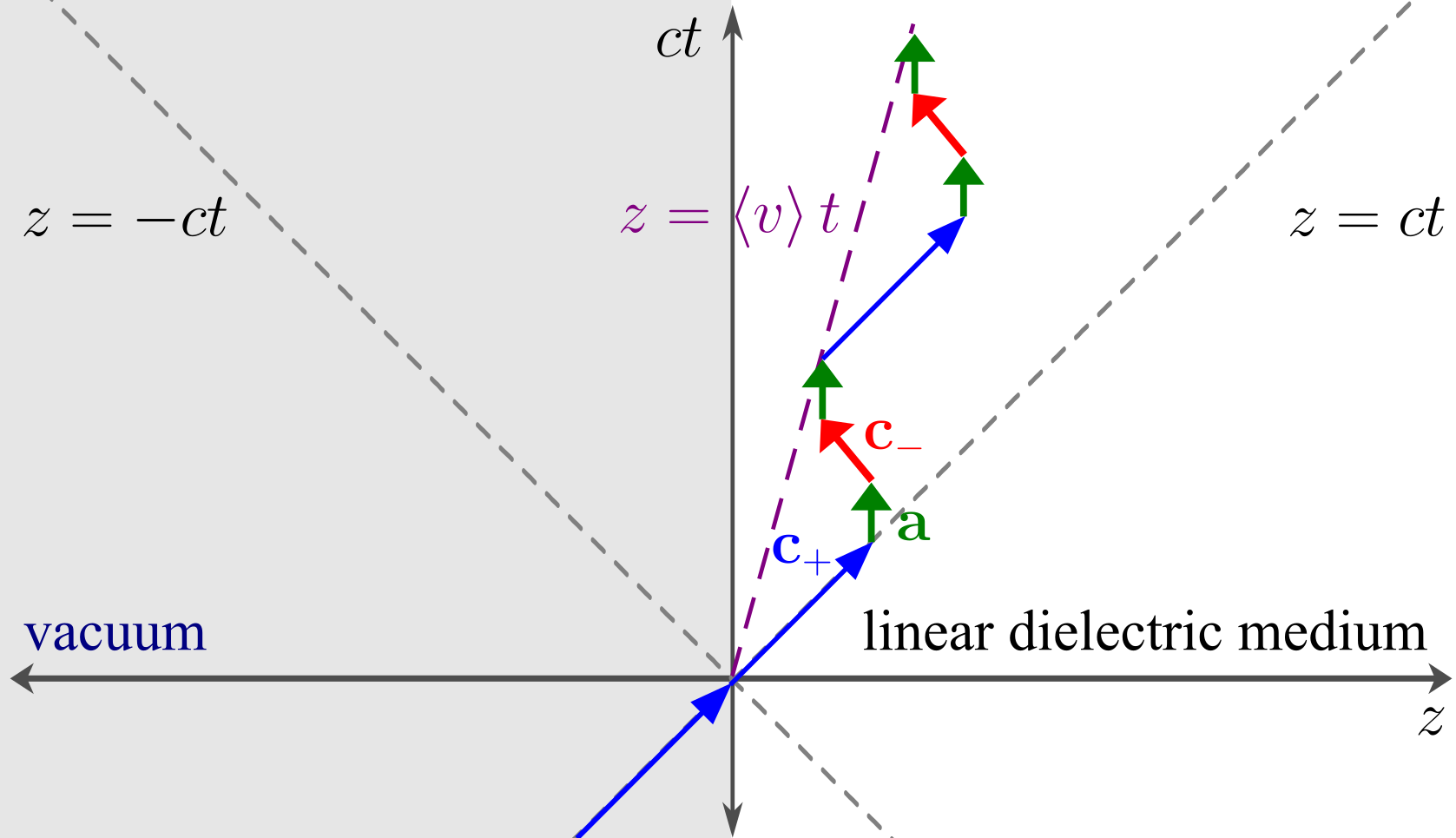}
\caption{Minkowski diagram illustrating the \textit{zitterbewegung} of a EM wave in a linear dielectric medium due to indirect coupling between right-handed ($\mathbf{c}_+$) and left-handed ($\mathbf{c}_-$) EM waves mediated by dipole oscillation ($\mathbf{a}$).}
\label{FigMinkowski}
\end{figure}

The velocity of EM wave operator is obtained as the momentum derivative of the Hamiltonian,
\begin{align}
\hat{\mathbf{v}} = \frac{\partial H}{\partial \mathbf{p}} = \left[
\begin{array}{cccc}
c \mathbf{s} & 0 & 0 & 0 \\
0 & -c \mathbf{s} & 0 & 0 \\
0 & 0 & 0 & 0 \\
0 & 0 & 0 & 0
\end{array}
\right].
\end{align}
The spin dynamics satisfies
\begin{align}
\frac{d \mathbf{s}}{dt} = \frac{1}{i \hbar} [\mathbf{s}, H] = \frac{1}{\hbar} \hat{\mathbf{p}} \times \hat{\mathbf{v}}, \label{EqConservedJ}
\end{align}
ensuring conservation of total angular momentum $d(\hbar \mathbf{s} + \hat{\mathbf{r}} \times \hat{\mathbf{p}})/dt = 0$. Equation~\ref{EqConservedJ} implies $d(\hat{\mathbf{p}} \cdot \mathbf{s})/dt = 0$, so $s_z$ is a good quantum number when $\hat{z} \parallel \hat{\mathbf{p}}$. Hence, the eigenvalue of spin along $\hat{z} \parallel \hat{\mathbf{p}}$ determines the observed velocity of EM wave: right-handed (left-handed) velocity of EM wave is parallel (anti-parallel) to $s_z$. The superposition of right- and left-handed EM waves, indirectly coupled via dipole oscillations, determines the EM wave's velocity in the dielectric medium, as shown in Fig.~\ref{FigMinkowski}:
\begin{align}
\langle \hat{\mathbf{v}} \rangle = \Psi^\dagger \hat{\mathbf{v}} \Psi = c ( \mathbf{c}_+^\dagger \mathbf{s} \mathbf{c}_+ - \mathbf{c}_-^\dagger \mathbf{s} \mathbf{c}_- ). \label{EqVelocity}
\end{align}

This phenomenon resembles the \textit{zitterbewegung} of electrons governed by the Dirac equation \cite{Pal2011}, except that the coupling between right- and left-handed EM waves here is mediated indirectly by dipole oscillations.

\subsection{Minkowski momentum from diagonalization} 
\begin{figure}[tb]
\includegraphics[width=\columnwidth]{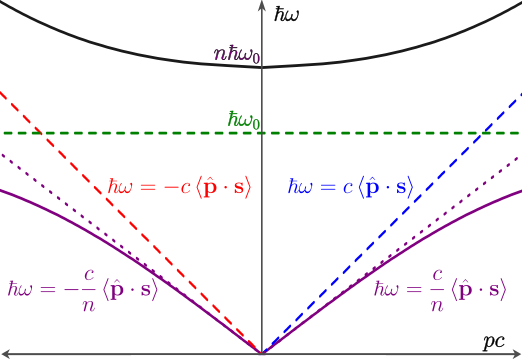}
\caption{Dispersion in a linear dielectric medium. Hybridization of right-handed (blue dashed line), left-handed EM waves (red dashed line), and dipole (green dashed line) modes results in linear dispersion $\hbar \omega = \pm p c / n$ (purple solid lines) at low frequencies.}
\label{FigDispersion}
\end{figure}

For small momenta $p c \ll \hbar \omega_0$, the Hamiltonian can be diagonalized as
\begin{align}
H = U \left[
\begin{array}{cccc}
\frac{c}{n} \mathbf{s} \cdot \hat{\mathbf{p}} & 0 & 0 & 0 \\
0 & -\frac{c}{n} \mathbf{s} \cdot \hat{\mathbf{p}} & 0 & 0 \\
0 & 0 & n \hbar \omega_0 & 0 \\
0 & 0 & 0 & -n \hbar \omega_0
\end{array}
\right] U^t, \label{EqDiagonal}
\end{align}
with
\[
U = \frac{\left[
\begin{array}{cccc}
1+n & 1-n & -\sqrt{n^2 - 1} & -\sqrt{n^2 - 1} \\
1-n & 1+n & -\sqrt{n^2 - 1} & -\sqrt{n^2 - 1} \\
-\sqrt{n^2 - 1} & -\sqrt{n^2 - 1} & -n - 1 & n - 1 \\
-\sqrt{n^2 - 1} & -\sqrt{n^2 - 1} & n - 1 & -n - 1
\end{array}
\right]}{2 n}
.
\]
Here $n$ is the refractive index from Eq.~\ref{EqRefractIndex}. The dispersion relations, corresponding to eigenfrequencies of the coupled system, are shown in Fig.~\ref{FigDispersion}. The first element $\hat{H}_{11}$ of the diagonalized matrix corresponds to the EM wave dispersion in the linear medium, yielding the Minkowski momentum 
\begin{equation}
p_M=\left<\hat{\textbf{p}}\cdot \textbf{s}\right>=\frac{n}{c}\langle\hat{H}_{11}\rangle,
\label{EqMinkowskiMomentum}
\end{equation}
where $\left<\hat{\textbf{p}}\cdot \textbf{s}\right>$ is illustrated by dispersion in Fig.~\ref{FigDispersion}. 

 The change in the momentum operator during refraction from vacuum to the linear dielectric medium conserves the tangential momentum $\hat{\mathbf{p}}_t$ at the refraction surface, since

\[
\frac{d \mathbf{p}}{dt} = -\nabla H.
\]

\begin{figure}
\includegraphics[width=\columnwidth]{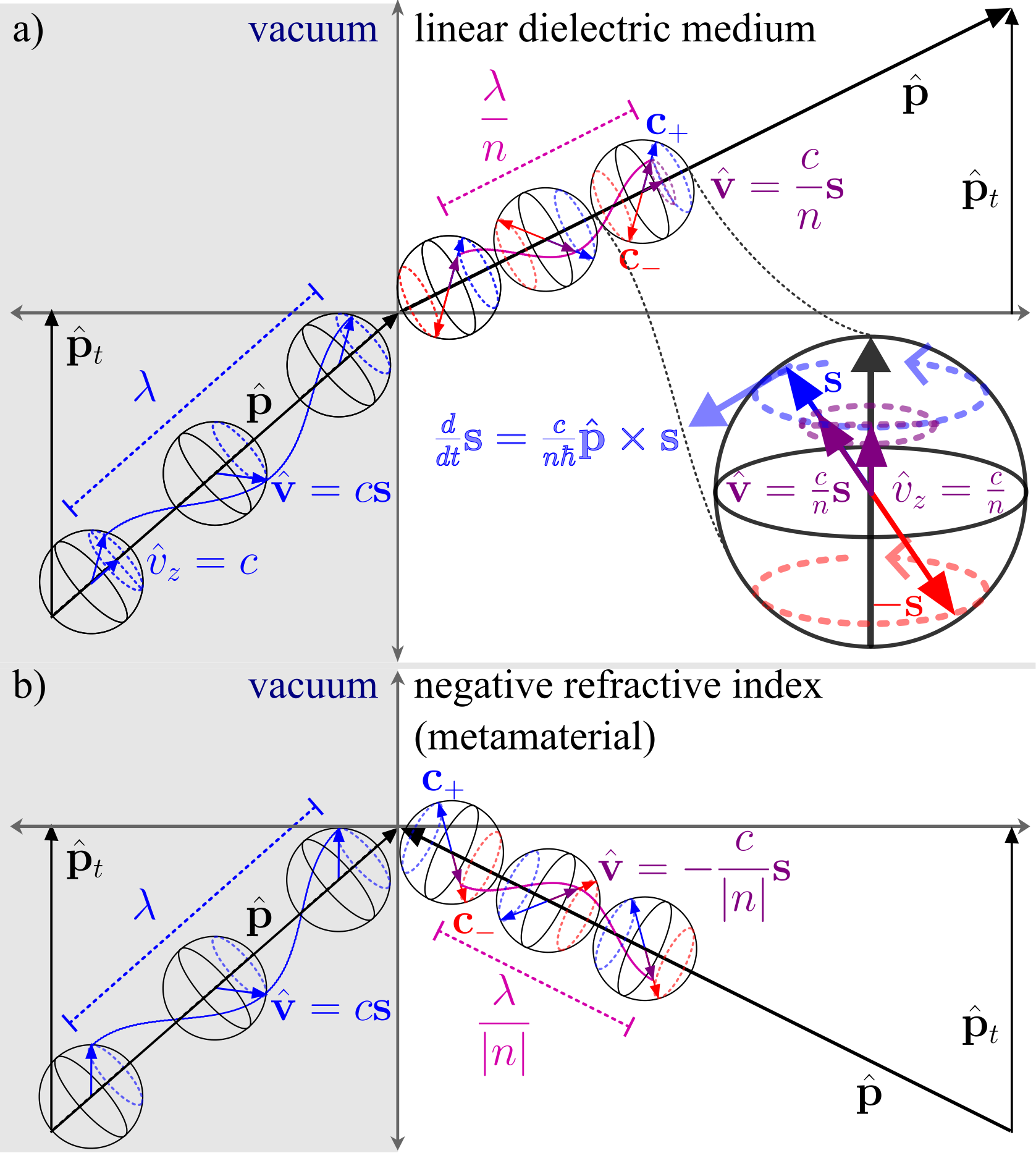}
\caption{(a) Refraction conserves tangential momentum $\hat{\mathbf{p}}_t$ at the interface. The velocity of the EM wave $\hat{\mathbf{v}}$ in the medium is a superposition of right-handed ($\mathbf{c}_+$) and left-handed ($\mathbf{c}_-$) EM waves, whose velocities are parallel and anti-parallel to the spin $\mathbf{s}$, respectively. The spin $\mathbf{s}$ precesses around $\hat{\mathbf{p}}$ with $s_z \parallel \hat{\mathbf{p}}$. (b) In metamaterials, the left-handed EM wave $\mathbf{c}_-$ contribution dominates, resulting in velocity opposite to $\hat{\mathbf{p}}$.}
\label{FigRefraction}
\end{figure}

The eigenvector $U_1$ associated with the first eigenvalue determines the contributions of $\mathbf{c}_+$ and $\mathbf{c}_-$ in the observed velocity of EM wave:
\[
\langle \hat{\mathbf{v}} \rangle_1 = U_1^t \hat{\mathbf{v}} U_1 = \mathbf{s} \frac{c}{n}.
\]
The relation between velocity and spin governs spin precession:
\begin{align}
\frac{d \mathbf{s}}{dt} = \frac{1}{\hbar} \hat{\mathbf{p}} \times \langle \hat{\mathbf{v}} \rangle_1 = \frac{c}{n \hbar} \hat{\mathbf{p}} \times \mathbf{s}.
\end{align}
The spin eigenvalue $s_z = 1$ along $\hat{z} \parallel \hat{\mathbf{p}}$ determines the velocity of EM wave $c/n$ in the medium. In metamaterials with negative refractive index, the left-handed EM wave contribution dominates, reversing the direction of velocity relative to $\mathbf{s}$ and $\hat{\mathbf{p}}$ (Fig.~\ref{FigRefraction}).

The transverse spin components $s_x$ and $s_y$ determine the EM wave phase, as illustrated in Fig.~\ref{FigRefraction}. Since the momentum $\hat{p}$ in the dielectric medium is $n$ times larger than in vacuum, the precession frequency remains unchanged. However, because the speed is reduced by $n$, the wavelength of the EM wave in the medium is correspondingly shorter by the same factor.

\subsection{Abraham momentum from acceleration}  
Thus far, we have identified conserved quantities of the EM wave in the medium: the momentum operator $\hat{p}$, the spin component $s_z$ parallel to it, and energy. The EM wave energy density entering the medium equals the sum of electromagnetic and dipole oscillation energy densities,
\begin{equation}
\frac{\hbar \omega}{V} = \frac{\varepsilon_0}{2} \left( E^2 + c^2 B^2 \right) + \frac{m}{2V} \left( \omega_0^2 r_e^2 + \dot{r}_e^2 \right).
\end{equation} 
This relation links acceleration and momentum change. The acceleration operator satisfies
\begin{small}
\begin{align}
&\frac{d \hat{\mathbf{v}}}{dt} =  \frac{c^2}{\hbar} \hat{\mathbf{p}} \times 
\left[
\begin{array}{cccc}
\mathbf{s} & 0 & 0 & 0 \\
0 & \mathbf{s} & 0 & 0 \\
0 & 0 & 0 & 0 \\
0 & 0 & 0 & 0
\end{array}
\right] + \frac{c\omega_p}{2i}
\left[
\begin{array}{cccc}
0 & 0 & \mathbf{s} & -\mathbf{s} \\
0 & 0 & -\mathbf{s} & \mathbf{s} \\
-\mathbf{s} & \mathbf{s} & 0 & 0 \\
\mathbf{s} & -\mathbf{s} & 0 & 0
\end{array}
\right], \notag \\
&\frac{d \langle \hat{\mathbf{v}} \rangle}{dt}  
= \frac{e \dot{\mathbf{r}}_e \times c^2 \mathbf{B}}{\frac{V \varepsilon_0}{2} \left(E^2 + c^2 B^2\right) + \frac{1}{2} \left(m \omega_0^2 r_e^2 + m \dot{r}_e^2\right)}.
\end{align}
\end{small}
Here the expectation value of the first term is zero because $\left<\textbf{s}\right>\parallel\hat{\textbf{p}}$. This relates acceleration to the reaction force on the dipole Lorentz force in Eq.~\ref{EqMomentum}:
\begin{align}
\frac{d}{dt} \frac{\hbar \omega}{c^2} \langle \hat{\mathbf{v}} \rangle = e \dot{\mathbf{r}}_e \times \mathbf{B}. \label{EqVdot}
\end{align}
We conclude that the quantity $\hbar \omega \langle \hat{\mathbf{v}} \rangle / c^2$ governs momentum transfer to the dipole. Its magnitude matches the Abraham momentum of a EM wave in a linear medium:
\[
\textbf{p}_A=\frac{\hbar \omega}{c^2} \langle \hat{\textbf{v}} \rangle .
\]

The acceleration can also be evaluated from the anticommutation relation,
\begin{small}
\begin{align}
&i\hbar\frac{d\hat{\mathbf{v}}}{dt}
=2\hat{\mathbf{v}}H-\left\{\hat{\mathbf{v}},H\right\}\notag\\
&=2c^2\left(\frac{\hat{\mathbf{v}}H}{c^2}-\left[
\begin{array}{cccc}
\frac{\{\hat{\textbf{p}}\cdot\mathbf{s},\mathbf{s}\}}{2} & 0 & \frac{\hbar\omega_p}{4c}\mathbf{s} & -\frac{\hbar\omega_p}{4c}\mathbf{s} \\
0 & \frac{\{\hat{\textbf{p}}\cdot\mathbf{s},\mathbf{s}\}}{2} & -\frac{\hbar\omega_p}{4c}\mathbf{s} & \frac{\hbar\omega_p}{4c}\mathbf{s} \\
\frac{\hbar\omega_p}{4c}\mathbf{s} & -\frac{\hbar\omega_p}{4c}\mathbf{s} & 0 & 0 \\
-\frac{\hbar\omega_p}{4c}\mathbf{s} & \frac{\hbar\omega_p}{4c}\mathbf{s} & 0 & 0
\end{array}
\right]\right). \label{EqAcceleration}
\end{align}
\end{small}
Taking the expectation value of Eq.~\ref{EqAcceleration} yields additional terms beyond Eq.~\ref{EqVdot}: 
\begin{align}
&\frac{d}{dt} \frac{\hbar \omega}{c^2} \langle \hat{\mathbf{v}} \rangle 
= -i2\omega\left(\frac{\hbar \omega}{c^2} \langle \hat{\mathbf{v}} \rangle - \left< \hat{\textbf{p}}_\textbf{s}\right>\right) \notag\\
&= -i2\omega\left(\frac{\hbar \omega}{c^2} \langle \hat{\mathbf{v}} \rangle-\left<\hat{\textbf{v}}\frac{i\hbar}{c^2}\frac{\partial}{\partial t}\right>\right)+e \dot{\mathbf{r}}_e \times \mathbf{B}. \label{EqForce}
\end{align}
Here $\hat{\mathbf{p}}_\mathbf{s}$ denotes the component of $\hat{\mathbf{p}}$ parallel to the spin.  
\begin{align*}
\hat{\textbf{p}}_\textbf{s}=\left[
\begin{array}{cccc}
\frac{1}{2}\{\hat{\textbf{p}}\cdot\mathbf{s},\mathbf{s}\} & 0 & 0 & 0 \\
0 & \frac{1}{2}\{\hat{\textbf{p}}\cdot\mathbf{s},\mathbf{s}\} & 0 & 0 \\
0 & 0 & 0 & 0 \\
0 & 0 & 0 & 0
\end{array}
\right].
\end{align*}
The additional term in Eq.~\ref{EqForce} produces a \textit{zitterbewegung}-type oscillation at frequency $2\omega$.  
In this framework, the Abraham momentum arises naturally as the time-averaged value of $\langle \hat{\mathbf{v}}i\hbar\partial_t \rangle / c^2$.

\section{Discussion}
\label{Sec_Discussion}
The novelty of our model lies in unifying spin dynamics, momentum transfer, and polarization precession within a semiclassical first-quantization framework.  
The two forms of electromagnetic momentum arise from different aspects of spin-projected operators.  
The Abraham momentum is related to the expectation value of the spin-projected momentum vector $\{\hat{\textbf{p}}\cdot\mathbf{s},\mathbf{s}\}/2$, directly linking it to the Lorentz-force reaction on the medium (Eq.~\ref{EqForce}).  
In contrast, the Minkowski momentum corresponds to the magnitude $p_M=\langle \hat{\mathbf{p}}\cdot\mathbf{s} \rangle$ (Eq.~\ref{EqMinkowskiMomentum}), obtained from the dispersion relation of the diagonalized Hamiltonian.  
This coexistence is reminiscent of the velocity of a Dirac relativistic electron: its magnitude equals the speed of light, yet its expectation value is reduced due to \textit{zitterbewegung}.  
This operator-level distinction provides a unified microscopic interpretation of how both momenta coexist and fulfill complementary physical roles in dielectric media.

This supports prior interpretations while providing a microscopic dynamical basis absent from field-theoretic formulations. Prior treatments such as in Ref.~\cite{Barnett2010} interpret the total electromagnetic momentum via energy-momentum tensor decomposition. There, the Abraham momentum corresponds to the kinetic part imparted to the medium, while the Minkowski momentum represents the canonical momentum of the light-matter system. Our model agrees with this interpretation but derives these quantities from the expectation value of the velocity and eigenvalue of Hamiltonian, without invoking macroscopic stress tensors.

In Ref.~\cite{Partanen2017} the photon mass drag and light-matter momentum transfer are computed using continuum field theory and relativistic energy-momentum conservation. While the resulting Abraham momentum is consistent with ours, their approach does not consider the spin dynamics of the EM wave. Our Hamiltonian allows one to track the coupled evolution of spin, velocity and linear momentum simultaneously, yielding a natural operator-level distinction between the two momenta.

In Ref.~\cite{Bliokh2017} the spin and orbital angular momentum (OAM) of light are investigated using canonical field decomposition. While their work thoroughly characterizes the transfer and conservation of optical spin and OAM, they embed the effect of the medium on the frequency-dependent permittivity $\varepsilon(\omega)$. In contrast, our model explicitly includes dipole degrees of freedom without absorbing these effects into an effective bulk parameter. This permits a more detailed examination of how polarization mediates the coupling between right- and left-handed EM modes. 

The coupled dynamics of the EM wave modes in our model closely resemble the \textit{zitterbewegung} predicted for relativistic electrons in the Dirac equation. Recent experimental work supports the physical plausibility of \textit{zitterbewegung}-like dynamics in photonic systems. Reference~\cite{Lovett2023} reported transverse oscillations in the position of light wavepackets confined in planar and lattice microcavities. These oscillations were interpreted as a manifestation of \textit{zitterbewegung}, arising from interference between two polarization-dependent optical modes. Although the experiment did not measure helicity or spin directly, the internal interference mechanism provides a useful analogy to the \textit{zitterbewegung} predicted in our model. This lends conceptual support to the idea that the wave-packet dynamics governed by coupled internal degrees of freedom can lead to observable oscillatory effects in optical systems.

\section{Conclusion}
\label{Sec_Conclusion}

In summary, we have analyzed the EM wave dynamics in a dielectric medium through the coupled Hamiltonian of EM waves and dipoles. The dipole oscillation in the medium mediates an indirect coupling between right-handed and left-handed EM waves, whose velocities are parallel and antiparallel to the spin $\mathbf{s}$, respectively. This helicity mixing is similar to the \textit{zitterbewegung} effect known in relativistic quantum mechanics, where internal degrees of freedom generate rapid oscillatory motion. The single-particle picture of \textit{zitterbewegung} reinforces the interpretation of light as possessing particle-like behavior, even in a classical-wave framework.

Conservation of spin ensures that the superposition of these EM waves reduces the velocity of EM wave in the medium, with $s_z$ aligned parallel to the momentum operator $\hat{\mathbf{p}}$. The precession of transverse spin components $s_{x,y}$ around $\hat{\mathbf{p}}$ determines the EM wave phase. Refraction into the dielectric medium conserves the precession frequency and the tangential momentum $\hat{\mathbf{p}}_t$, thereby increasing the magnitude of $\hat{\mathbf{p}}$. This increase corresponds to the refractive index and gives rise to the Minkowski momentum, which determines the direction of propagation according to Snell's law.

In contrast, the Abraham momentum represents the actual transfer of momentum to the medium.  
In our formulation, it is obtained as the expectation value of the spin-projected momentum vector and appears in the acceleration equation as the term balancing the Lorentz-force reaction on the dipoles.  
The Minkowski momentum, on the other hand, corresponds to the magnitude of the spin-projected momentum and governs canonical wave propagation.  
This coexistence mirrors the case of a relativistic Dirac electron, whose velocity magnitude equals the speed of light, while its expectation value is reduced by \textit{zitterbewegung}. 
We also predict a \textit{zitterbewegung}-like oscillation due to helicity mixing, which emerges from the dipole-mediated coupling of polarization modes. While this effect does not require a quantized field, it reflects internal dynamics typically associated with particle behavior, offering a conceptual bridge between classical wave motion and the particle picture. Our results clarify the coexistence of both Abraham and Minkowski momenta in dielectric media, assigning distinct physical roles to each.

\section*{Acknowledgment} 
We acknowledge supports from Universitas Indonesia via PUTI Research Grant No. PKS-182/UN2.RST/HKP.05.00/2025

%

\end{document}